\documentclass[pra,aps,showpacs,superscriptaddress,twocolumn,notitlepage]{revtex4-2}

\newcommand{\beq}{\begin{equation}}
\newcommand{\eeq}{\end{equation}}
\newcommand{\beqa}{\begin{eqnarray}}
\newcommand{\eeqa}{\end{eqnarray}}

\usepackage{amsmath,amsfonts,amssymb,amsthm,graphics,graphicx,epsfig,bbm}
\usepackage[colorlinks=true,citecolor=red,linkcolor=blue,urlcolor=blue]{hyperref}
\usepackage[usenames]{color}
\usepackage{graphicx}
\usepackage{subfigure}
\usepackage{amsmath}
\usepackage{epsfig}
\usepackage{dcolumn}
\usepackage{bm}
\usepackage{color}
\usepackage{times}
\usepackage{epstopdf}
\usepackage{amssymb}
\usepackage{amstext}
\usepackage{latexsym}
\usepackage{hyperref}
\usepackage{amsfonts}
\usepackage{psfrag}
\usepackage{soul,xcolor}
\usepackage[normalem]{ulem}
\usepackage{dsfont}
\usepackage{adjustbox}

\begin{document}

\title{Time-Optimal Quantum Driving by Variational Circuit Learning}

\author{Tangyou Huang}
\affiliation{Department of Physical Chemistry, University of the Basque Country UPV/EHU, Apartado 644, 48080 Bilbao, Spain}
\affiliation{International Center of Quantum Artificial Intelligence for Science and Technology (QuArtist) \\ and Department of Physics, Shanghai University, 200444 Shanghai, China}

\author{Yongcheng Ding}
\affiliation{Department of Physical Chemistry, University of the Basque Country UPV/EHU, Apartado 644, 48080 Bilbao, Spain}

\author{L\'eonce Dupays}
\affiliation{Department  of  Physics  and  Materials  Science,  University  of  Luxembourg,  L-1511  Luxembourg, Luxembourg}

\author{Yue Ban}
\affiliation{TECNALIA, Basque Research and Technology Alliance (BRTA), 48160 Derio, Spain}

\author{Man-Hong Yung}
\affiliation{Central Research Institute, 2012 Labs, Huawei Technologies, Shenzhen 518129, China}
\affiliation{Department of Physics, Southern University of Science and Technology, Shenzhen 518055, China}
\affiliation{Shenzhen Institute for Quantum Science and Engineering, Southern University of Science and Technology, Shenzhen 518055, China}

\author{Adolfo del Campo}
\affiliation{Department  of  Physics  and  Materials  Science,  University  of  Luxembourg,  L-1511  Luxembourg, Luxembourg}
\affiliation{Donostia International Physics Center,  E-20018 San Sebasti\'an, Spain}

\author{Xi Chen}
\affiliation{Department of Physical Chemistry, University of the Basque Country UPV/EHU, Apartado 644, 48080 Bilbao, Spain}
\affiliation{EHU Quantum Center, University of the Basque Country UPV/EHU, Barrio Sarriena, s/n, 48940 Leioa, Spain}

\date{\today }
	
\begin{abstract}	

The simulation of quantum dynamics on a digital quantum computer with parameterized circuits has widespread applications in fundamental and applied physics and chemistry. In this context, using the hybrid quantum-classical algorithm, combining classical optimizers and quantum computers, is a competitive strategy for solving specific problems. We put forward its use for optimal quantum control. We simulate the wave-packet expansion of a trapped quantum particle on a quantum device with a finite number of qubits. We then use circuit learning based on gradient descent to work out the intrinsic connection between the control phase transition and the quantum speed limit imposed by unitary dynamics.
We further discuss the robustness of our method against errors
and demonstrate the absence of barren plateaus in the circuit. The combination of digital quantum simulation and hybrid circuit learning opens up new prospects for quantum optimal control. 

\end{abstract}
	
\maketitle
	
\section{Introduction}
Following the vision by Feynman \cite{feynman}, quantum simulation has acquired a potentially-disruptive role in the development of contemporary science and technology, given the prospects of harnessing the advantage of using a quantum computer (QC) for specific applications. In recent decades, quantum simulation has been used to probe the dynamics of condensed matter systems \cite{CondensateScience2011,CondensateSolano2015prx}, for quantum chemistry \cite{qChemistry2018prx, qChemistry2016prx}, and as a test-bed for nonequilibrium statistical mechanics, e.g., in studying thermalization and nonequilibrium behavior of many-body systems \cite{therm2o11prl,Nonequilibrium2011RMP}. Quantum simulation is also expected to impact high-energy physics,
given the potential to facilitate the study of lattice gauge theories \cite{Banuls2020} and gauge-gravity duality \cite{GarciaALvarez17,Luo2019}, among other examples. 

The use of a digital quantum simulator (DQS) based on the gate model offers a prominent approach in current Noisy Intermediate-Scale Quantum (NISQ) devices \cite{NISQ} and has gained relevance with theoretical and experimental progress  \cite{DQS2015prx,Nature2016gauge,Zoller2019nature}.
In particular, DQS can be used to implement
variational quantum algorithms (VQAs), under development for quantum optimization \cite{qaoa}, quantum machine learning \cite{DQML}, and quantum control \cite{DQcontrol}. Their formulation generally approximates the continuous time evolution by discrete, finite Trotter steps \cite{Trotter1,suzuki1993general,Trotter2,Trotter3} implemented by a sequence of quantum gates, with controlled accuracy, in principle. However, balancing the number of Trotter steps and imperfections of quantum circuits in experiments is still a fundamental challenge. In this sense, various optimization scenarios aim at quantum error mitigation  \cite{GAs2016prl,ErrorCorrection2019prl,TrotterOptim2020pra} for achieving a good precision of the quantum simulation with limited quantum resources. Among those, the machine-learning-enhanced optimization protocol \cite{heyl2021prl,RL2018prx,RL2018prb,RL2018pra} utilizes a feedback loop between the quantum device and a classical optimizer. This approach is particularly useful in the field of hybrid quantum algorithms \cite{VQE1,VQE2} with current quantum hardware. Nonetheless, solving the quantum optimal control problem by VQAs in a NISQ device is still an open challenge \cite{DQcontrol}.

In this work, we propose a circuit learning scheme based on gradient-descent (GD) for time-optimal quantum control. As a concrete example, we consider a quantum particle trapped in time-varying parabolic potential. We use a qubit register and encode the spatial wave function using the basis of $n$-qubit states. We then reproduce the exact state evolution on a designed quantum circuit using a digital algorithm \cite{quadraticOperator,Molecular2020prl}. We optimize the control function to achieve maximum-fidelity control by using the GD-based circuit learning. We further unveil the connection between a control phase transition and the quantum speed limit, i.e., the minimum time for a quantum state to evolve into a distinguishable state under a given dynamics. We demonstrate that the fidelity-based GD method avoids a large number of measurements by comparison to the reinforcement learning protocol \cite{CPT2018prx} and show how it can be accelerated by choosing different quantum quantities as the cost function.

In the following two sections, we introduce the quantum algorithm for the circuit realization of a quadratic Hamiltonian and discuss the time-dependent harmonic oscillator as an example. We then explore a fidelity-based GD method for maximum-fidelity control in a nonequilibrium expansion process and characterize the efficiency through various cost functions.
The relation of the quantum speed limit to the control phase transition is then discussed. Finally, we establish the fault tolerance of our method against the quantum errors in experiments and also address the problem of barren plateaus in the parameterized circuit.

\section{PRELIMINARIES AND NOTATION}

\subsection{Time-dependent quantum harmonic oscillator}

We exemplify our approach by considering the time-dependent harmonic oscillator (TDHO), described by the Hamiltonian 
\beq
H(t) = \frac{\hat{p}^2}{2m} +\frac{1}{2}m \omega^2(t)[\hat{x}-x_0(t)]^2,
\eeq
where $\omega(t)$ and $x_0(t)$ are tunable and represent the trap frequency and the location of the trap center, respectively. The TDHO is an ideal model for benchmarking quantum control algorithms since its dynamics admits exact closed-form solutions. In particular, we focus on the case with $x_0(t)=0$ and look for the expansion of the wave packet induced by a modulation of the trap frequency $\omega(t)$ from an initial value $\omega_0$ to a final one $\omega_f$. This model has many applications, including the cooling of a particle in an optical trap \cite{chenprl104}, mechanical resonators \cite{LianAoPRA}, and tunable transmon superconducting qubits \cite{JuangoPRAppl}. The ground state of $H(0)$ evolves into the time-dependent Gaussian state \cite{chenprl104}
\beqa
\label{wavefunction}
\Psi(t,x) = \left(\frac{m\omega_0}{\pi\hbar b^2}\right)^{1/4}\exp\left[-\frac{i}{2} \int_0^t \frac{\hbar\omega_0}{b^2}dt'\right]\nonumber \\ 
\times\exp\left[\frac{im}{2\hbar}\left(\frac{\dot{b}}{b}+i\frac{\omega_0}{b^2}\right)x^2\right],
\eeqa
where the time-dependent scaling factor $b(t)>0$ characterizes the width of the wave packet and satisfies the auxiliary equation 
\beqa\label{ermakov}
\ddot{b}+\omega^2(t) b = \frac{\omega_0^2}{b^3}.
\eeqa
A primary numerical solver of quantum dynamics is the so-called \textit{split-operator method} (SOM), also known as the split time propagation scheme \cite{Kosloff1988}. For the sake of convenience, one usually sets dimensionless variables based on physical units of energy $\epsilon= \hbar \omega_0$,  length $b_{\text{HO}}= \sqrt{\hbar/m\omega_0}$, and time $\tau=1/\omega_0$. In a classical computer, one defines a $N$-dimensional vector as $\Psi(\textbf{r})$ for encoding the amplitude of the wave function on the space grid $\textbf{r}=[x_0,~x_1,\cdots,x_{N-1}]$. Note that the kinetic energy operator $\hat{T}=\hat{p}^2/2$ and the potential operator $\hat{V}=\omega^2(t)\hat{x}^2/2$ do not commute. Thus, the following approximation stands for small $dt$ with an error  $\mathcal{O}(dt^3)$
\beq
e^{-iHdt} \approx e^{-\frac{i}{2}\hat{V}dt}e^{-i\hat{T}dt}e^{-\frac{i}{2}\hat{V}dt},
\eeq
evolving the wave function in a Trotter step. A common trick in implementing this method uses forward and inverse Fourier transforms to change the representation of the quantum state between the real space $\textbf{r}$ and momentum space $\textbf{k}$, in which the kinetic energy operator becomes diagonal in $\textbf{k}$,  simplifying the numerical calculation.

\subsection{Time-optimal control}
The frictionless expansion of quantum particles trapped in a time-varying harmonic trap can be formulated as a time-optimal control problem by minimizing the time of the process $t_f$ \cite{chenprl104,stefanatos2010frictionless,hoffmann2011time,Chaos,Dupays21}. It follows from Pontryagin's maximum principle that the control Hamiltonian for all $t\in[0,t_f]$ takes the form \cite{stefanatos2010frictionless}
\begin{equation}
H_c[x_1,x_2,p_1,p_2] = p_1x_2 + \frac{p_2}{x_1^3} - p_2x_1u(t),
\end{equation}
where the state $x_1=b$,~$x_2=\dot{b}/\omega_0$, and the controller $u(t)=\omega^2(t)/\omega_0^2$ are governed by the Ermakov equation~\eqref{ermakov}. Here, $p_1$ and $p_2$ are the conjugate momentum of $x_1$ and $x_2$, respectively. Substituting the control Hamiltonian into the canonical equation leads to the cost equations 
\begin{eqnarray}
\dot{p_1}&=&\left(u+\frac{3}{x_1^4}\right)p_2,\\
\dot{p_2}&=&-p_1.
\end{eqnarray}
If the controller is bounded as $\delta_1\leq u(t) \leq\delta_2$, the time-optimal control has a bang-bang form, i.e., it is a piece-wise function and constant in each interval. For a specific problem with $b(0)=1$ and $b(t_f) = \sqrt{\omega_0/\omega_f}=\gamma$, consider the feasible three-jump protocol 
\beq\label{bangbang}
u(t) =
\begin{cases}
1~~~~~~~~~~~~~~~t=0\\
\delta_1 ~~~~~~~0<t \leq t_1\\
\delta_2~~~~~~~~t_1 <t < t_1+t_2\\
1/\gamma^4~~~~~t = t_f^{\text{opt}} = t_1+t_2
\end{cases},
\eeq
where the switching time $t_1$ and the optimal operation time $t_f^{\text{opt}} = t_1 + t_2$ can be calculated by integrating the Ermakov equation~\eqref{ermakov} by using boundary conditions. This yields the closed-form exact time-optimal driving protocol $u(t)$ with
\begin{eqnarray}
t_1 &=& \frac{1}{\sqrt{\delta_1}}\sinh^{-1}\sqrt{\frac{\delta_1(\gamma^2-1)(\gamma^2\delta_2-1)}{(\delta_1-\delta_2)\gamma^2(1-\delta_1)}}, \nonumber\\
t_2 &=& \frac{1}{\sqrt{\delta_2}} \sin^{-1}\sqrt{\frac{\delta_2(\gamma^2-1)(1-\gamma^2\delta_1)}{(\delta_1-\delta_2)(1-\gamma^4\delta_2)}}. 
\label{optimaltime}
\end{eqnarray}

	\begin{figure*}
	\centering
	\includegraphics[width=2\columnwidth]{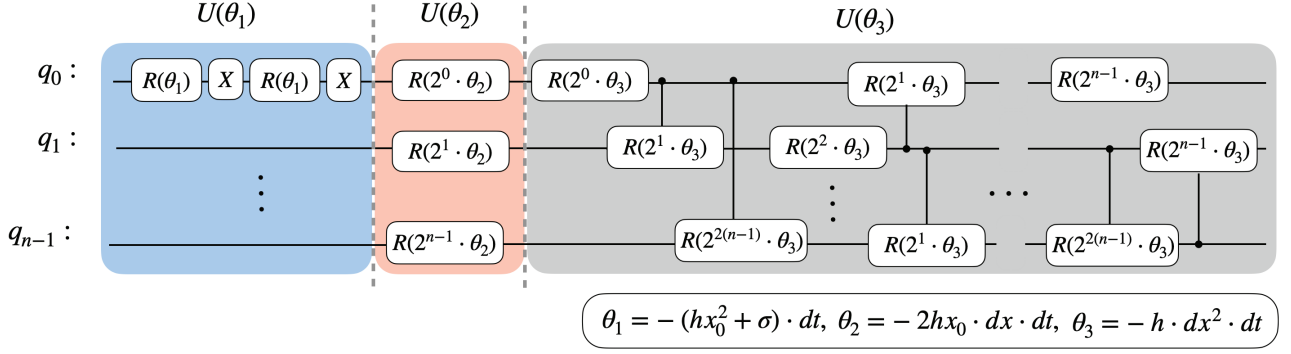}
	\caption{ The circuit realization of time-evolution operator for a quadratic Hamiltonian as defined by Eq. (\ref{quadratic operator}).
			The single phase gate and Pauli-X gate are schematically represented by the symbols $R(\cdot)$, $X$, and the corresponding control gates are shown by vertical lines with black circles.
	}\label{figure1}
\end{figure*}
	
\subsection{Quantum speed limit}

Quantum speed limits (QSLs) provide fundamental upper bounds on the speed of quantum evolution \cite{Deffner17rev}. They have wide-spread applications ranging from quantum metrology to optimal control \cite{qslwithOC,Funo17,Campbell17}. QSLs are formulated by choosing a notion of distance between quantum states and identifying a maximum speed of evolution. For isolated systems described by a time-independent Hamiltonian, two seminal results are known. The Mandelstam-Tamm QSL determines the maximum speed of evolution in terms of the energy dispersion \cite{MT45}, while the Margolus-Levitin bound uses the mean energy above the ground state instead \cite{ML98}. The interplay of these bounds has recently been demonstrated in a trapped system made of ultracold atoms that are suddenly quenched \cite{Ness21}. For a generic driven system,  only an analog of the Mandelstam-Tamm bound is known \cite{AA90,Uhlmann92,deffner2013energy}. 

Consider the quantum unitary dynamics generated by a time-dependent Hamiltonian according to the Schr\"odinger equation. The distance between the initial state and the time-dependent state in projective Hilbert space can be quantified by the Bures angle
\beq
\mathcal{L}(\psi_0,\psi_t)=\arccos( | \langle \psi_0 | \psi_t \rangle|)\in[0,\pi/2].
\eeq
The  minimum time scale required to sweep a given Bures angle is lower bounded by 
\beq\label{tau_QSL}
\tau _{\rm QSL} =  \frac{1}{ \overline{\Delta E} }\mathcal{L}(\psi_0,\psi_t),
\eeq
where the speed of evolution is set by the time-averaged energy dispersion
\beq
\overline{\Delta E} = \frac{1}{t}\int_0^tds\sqrt{\langle \psi_s |\hat{H}(s)^2|\psi_s\rangle-\langle \psi_s |\hat{H}(s)|\psi_s\rangle^2}.
\eeq
The QSL $\tau_{\rm QSL}$ is thus approached by maximizing the energy dispersion at all times. In control protocols for wave-packet expansion, we consider the evolution of the ground state of the trap with initial trapping frequency $\omega_0$  to the ground state of the trap with final frequency $\omega_f$. Provided that the control protocol, specified by $u(t)$, has unit efficiency in preparing the target state, the Bures angle is fixed, and the corresponding QSL reads
\beq
\label{qsl-ho}
\tau_{\rm QSL} = \frac{\hbar}{\overline{\Delta E}}\sqrt{\frac{2 \gamma}{1 +\gamma^2}}.
\eeq
We note that $\tau_{\rm QSL}$ should be distinguished from the minimum time $t^{\rm opt}_f$ in the preceding time-optimal control with the trap frequency bounded.

	\subsection{ Fidelity susceptibility}
	
Generally, the final state $\rho_{f} =|\Psi_{f}(x)\rangle\langle\Psi_{f}(x)|$  upon completion of a control protocol at $t=t_f$ differs from the target state $\rho_{\rm tar} =|\Psi_{\rm tar}(x)\rangle\langle\Psi_{\rm tar}(x)|$ one wishes to prepare. 
Let us consider the fidelity $F$ between these two states 
\beq
\label{fidelity}
		F(\rho_{\rm tar},\rho_f)=\left[{\rm Tr}(\sqrt{\sqrt{\rho_{\text{tar}}}\rho_f\sqrt{\rho_{\text{tar}}}})\right]^2,
	\eeq
	where ${\rm Tr}(\cdot)$ denotes the trace operation.

The {\it fidelity susceptibility} $\chi_f$  quantifies the fidelity response to a slight change of driving parameter \cite{FS2007pre,FS2008prb,FS2010GSJ}. For a functional Hamiltonian $H(f_t)$ parameterized by $f_t$, let $|\Psi_0(f_t)\rangle$ be the ground state. We assume $f_t$ to be a function of time and consider a variation on the control function $f_t\to f_t+\delta f$, where $\delta f\to 0$ is small enough to apply perturbation theory. As a result, the perturbed ground state is $|\Psi_0(f +\delta f)\rangle$. The fidelity susceptibility,
without loss of generality, is defined as
\beq
\chi_f \equiv \frac{-2\ln (F_{\delta f})}{\delta f},
\eeq
where the fidelity $F_{\delta f}=F[\rho_{0}(f), \rho_0(f+\delta f)]$.  The \textit{fidelity susceptibility} quantifies the sensitivity of the fidelity to variations of the control functions. In other words, the fidelity susceptibility can be used as a cost function to accelerate the convergence of the optimization process. For the sake of simplicity, we assume that $\delta f\to 0$ is a time-independent real value.

	\section{Quantum circuit realization of quadratic Hamiltonian }

	Next, we present the algorithm for the circuit realization of quadratic  Hamiltonians using a finite set of elementary quantum gates. We focus on DQS of the continuous-variables system 
	and encode a wave packet onto a $n$-qubit register. Quantum states of this register can be described in binary notation
	\beq
	|\Phi\rangle = \sum^{2^n-1}_{i=0} c_i|i\rangle,
	\eeq
	using the computational basis $|i\rangle=|q_{n-1}\rangle\otimes \cdot\cdot\cdot \otimes|q_1 \rangle\otimes |q_0 \rangle$ with $q_0,q_1,\dots,q_{n-1}\in\{0,1\}$ , and the corresponding amplitudes $c_i$  normalized as $\sum^{2^n-1}_{i=0}|c_i|^2\equiv1$. To solve the time-dependent Schr\"odinger equation on a quantum computer with a $n$-qubit register, we discretize the continuous variables associated with the spatial coordinate $x$ and time $t$, and subsequently map the coordinate space $x$ into the Hilbert space of $n$ qubits.
	Specifically, the compact continuous spatial domain $x\in[-L,L]$ is approximated by a lattice of $2^n$ points spaced by a constant interval $dx = L/(2^{n-1}-1)$.
	A wave packet can be encoded in the state of the $n$-qubit register as
	\beqa\label{q_states}
	|\Phi\rangle &=& \sum^{2^n-1}_{i=0} \Psi(x_i)|i\rangle \\
	&=&\Psi(x_0)|0\cdots0\rangle+\dots+\Psi(x_{2^n-1})|1\cdots1\rangle,
	\eeqa
	which reproduces the vectorized wave function for the following quantum analog of the SOM in the Hilbert space. As in numerical discretization methods, this encoding of $\Psi(\textbf{r})$ provides, in principle, satisfactory accuracy when the lattice length $dx$ is much smaller than any characteristic length scale of the wave packet. The preparation of arbitrary initial qubit states $|\Phi\rangle$ based on the initialized wave-packet $\Psi(\textbf{r})$ can be approached by the variational quantum eigensolver (VQE)
	\beq\label{vqe}
	|\tilde{\Phi}\rangle = \prod_{i=1}^{p}\left[ \prod_{q=0}^{n-1}\left(U^{q,i}\right)U_{\text{ENT}}\right]|+\rangle^{\otimes n},
	\eeq
	where  $|+\rangle=\frac{1}{\sqrt{2}}(|0\rangle+|1\rangle)$ is a single-qubit state, the unitary $U^{q,i}(\theta)= R_z^{q,i}(\theta_1^{q,i})R_x^{q,i}(\theta_2^{q,i})R_z^{q,i}(\theta_3^{q,i})$ is a universal single-qubit gate, and $U_{\text{ENT}}$ represents CNOT gates that entangle the neighboring qubits with periodic boundary conditions. In this way, an approximated initial state $|\tilde{\Phi}\rangle$ is prepared by optimizing $3pn$ parameters to minimize the cost function.
	
	Next, consider the digital quantum simulation aimed at reproducing an equivalent SOM.
	The dynamics of the wave packet is described by  
	\beq
	\Phi(t+dt) \approx e^{-iH(t)dt}\Phi(t),
	\eeq
	where $e^{-iHdt}$ is the time-evolution operator for the time step $dt$.

	We use the quantum Fourier transform (QFT) as the quantum analog of the inverse discrete Fourier transform, which is key to the efficient implementation of the SOM. Hence, the wave function $|\Phi\rangle$ is evolved as
	\begin{equation}
	|\tilde{\Phi}(t+dt)\rangle = \mathcal{V}(t)_{dt/2}\text{QFT}\mathcal{T}(t)_{dt}\text{QFT}^\dag\mathcal{V}(t)_{dt/2}|\tilde{\Phi}(t)\rangle,
	\end{equation}
	where $\mathcal{V}(t)$ and $\mathcal{T}(t)$ are the potential operator and the kinetic-energy operator in the real space and momentum space, respectively. In other words, they are both quadratic, and their diagonal elements can be written as
	\begin{equation}\label{quadratic operator}
	\mathcal{A}_{jj} = \exp\left\{-i dt [h(j dx +x_0)^2+\sigma]\right\},
	\end{equation}
	where other off-diagonal elements are zero. The preliminary result in Ref. \cite{quadraticOperator} demonstrates that the quadratic Hamiltonian can be exactly decomposed into a quantum circuit. In Fig.~\ref{figure1}, we plot the quantum circuit for implementing a quadratic Hamiltonian in the computational basis for DQS. The decomposition is verified by the DQS of nonadiabatic processes in molecular systems \cite{Molecular2020prl}.

	\section{Fidelity-based Gradient Descent}

	\subsection{Initial state preparation}
\begin{figure}
		\centering
		\includegraphics[width=\columnwidth]{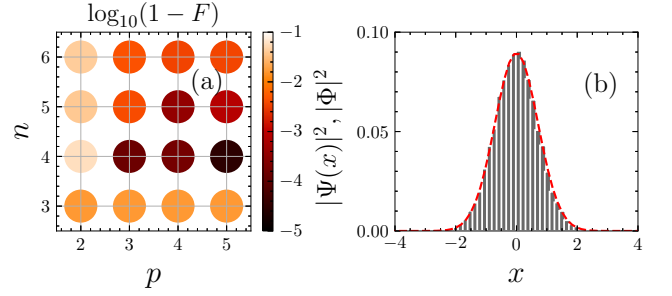}
		\caption{ The fidelity of states preparation by VQE as a function of qubit number $n$ and parameter depth $p$ in (a), and corresponding probability distribution of qubits $|\Phi|^2$ compared with density of wave function $|\Psi(x)|^2$  in (b), with fidelity $F = 0.996$ for $n = 6$ and $p=4$.
		}\label{figure2}
	\end{figure}

	The first step is to encode the information of the wave packet into the state of the qubit register. The accuracy of the preparation of a target state of qubits by VQE depends on the qubit number $n$ and the parameter depth $p$, see Eq. (\ref{vqe}). 
	As we know, the depth $p$ and qubit number $n$ exponentially increase the complexity of VQE. 
	In Fig. \ref{figure2}, we compare the fidelity of state preparation in the coefficient grid ${n,p}$ in (a).
	In (b), without loss of accuracy, we choose $n = 6$, $p = 4$, and present the resulting $n$-qubits states for the corresponding density (\ref{wavefunction}) with $\omega_0 = 1$.
	In what follows,  the numerical results are produced by the quantum simulator {\tt statevector simulator} on the {\tt qiskit} platform, which admits no errors, decoherence, and imperfections at all. We will consider the noise of an actual quantum device in the discussion.

	\subsection{The maximum-fidelity control}
	
\begin{figure}[t]
		\centering
		\includegraphics[width=\columnwidth]{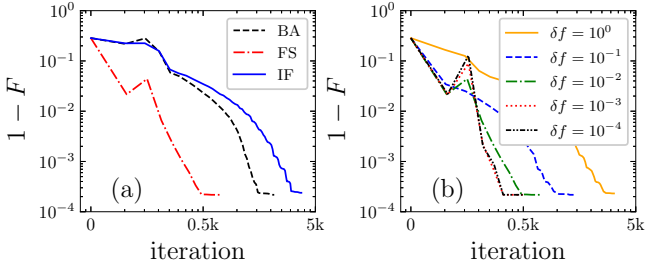}
		\caption{(a)  The infidelity $1-F$ as a function of the training iteration using three different loss functions: fidelity susceptibility (FS) with $\delta f=0.01$, infidelity (IF), and Bures angle (BA). (b)
				The infidelity as a function of the iteration step for different $\delta f$ is compared when the fidelity susceptibility (FS) is taken as a loss function.  Parameters: $N_t = 50$, $\omega_f = 0.1$, $n = 6$, $p = 4$, $\delta_1 = 10^{-6}$,  $\delta_2 = 1$, and $t_f = 3.152$.}  
		\label{figure3}
	\end{figure}

	A parametric optimization problem is usually mapped into the minimization of a given cost function to find its local minimum with the Gradient Descent (GD) algorithm.
	The optimal solution $M=\{m_0,m_1,...,m_i\}$ is obtained by minimizing the cost-value $c =J(M)$,  and can be expressed as
	\beq
	M^{\text{opt}} = \min_{c}J(M).
	\eeq
	In our case, the control function is the trap frequency $f(t) =\omega^2(t) $ that is piece-wise on the discrete time $t\in [0, t_f ]$ involving $N_t$-intervals.
	Accordingly, the control tuple $f(t) =\{f(0),f(dt),...,f(t_f)\}$ is constrained by $\delta_1\le|f(t)|\le\delta_2$, $|f(t+dt)-f(t)|\le\Delta f$ and the boundary conditions $f(0)=1$ and $f(t_f) = 0.01$ are considered for trap expansion with $\omega_f/\omega_0= 1/10$. Then, we optimize $\omega$-dependent parameters  $M = \{\theta_3(\omega)\}_t$ in the circuit shown in Fig. \ref{figure1} by minimizing the loss value which can be generated by the measurements on the qubits at $t = t_f$.
    Here, we exploit three different cost functions: the infidelity (IF) $(1-F)$, the fidelity susceptibility (FS)  $\chi_f$, and the Bures angle (BA) $\mathcal{L}(\psi_t,\psi_{\text{tar}})$.
    For simplicity, we 
	initialize the controller $f(t)$ with a linear dependence of the form $f(t)=(\omega_f^2-\omega_0^2)(t/t_f)+\omega_0^2$.
	This yields a parametric constrained minimization problem in this case.
	We exemplify the optimization process of finding maximum-fidelity policy for different cost functions in Fig. \ref{figure3} by using the optimizer $\tt{SLSQP}$ \cite{SLSQP} based on the $\tt{scipy}$ \cite{scipy} platform.
	To this end, we assign the total time $t_f = t_f^{\text{opt}}$ calculated in (\ref{optimaltime}), and use the GD method to minimize the cost function for obtaining the maximum-fidelity control function.
	In Fig. \ref{figure3}, we present infidelities versus the GD iteration when using each cost function.
	The control with maximum fidelity is obtained with their convergence.
	One can see that the learning rate of FS outperforms the others with the same optimizer in (a), and we compare the FS-based GD with various coefficients: $\delta f = [10^{-4}, 10^{-3},10^{-2},10^{-1},10^{0}]$ in (b).
	Because the greatest convergent rate of the optimization process arises when $\delta f \le 10^{-3}$, we employ the FS $\chi_f$ as the cost function with the coefficient $\delta f = 10^{-3}$ to design the maximum fidelity policy in the GD algorithm.
	It is worth emphasizing that reducing the training iteration is equivalent to decreasing the accumulation of operation errors.
    As a result, we can improve the accuracy of DQS with the same quantum volume.

\begin{figure}
		\centering
		\includegraphics[width=\columnwidth]{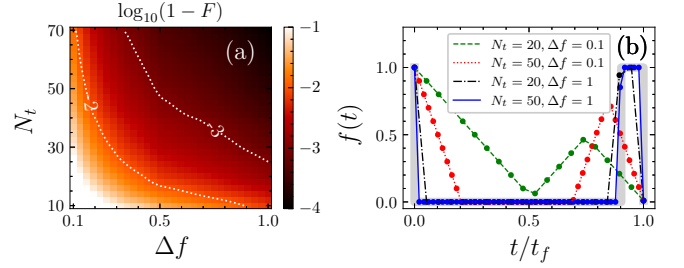}
		\caption{(a) Fidelity as function of $\Delta f$ and $N_t$ with total time $t_f=3.152$. The two dashed contour curves refer to $F = 0.99,0.999$ and are labeled by $-2,-3$, respectively.
				In (b), we present the trained fidelity-optimal controls compared to the bang-bang control (solid gray line). The fidelity takes values
				 $F = 0.84,0.98,0.998,0.9998$ in the four cases illustrated in the legend from top to bottom.
				Other parameters are chosen as in Fig. \ref{figure3}}.
		\label{figure4}
	\end{figure}

	By considering the trade-off between the complexity and accuracy of digitized circuits, we analyze the fidelity achieved by the maximum-fidelity policy for different numbers of Trotter steps $N_t$ and the constraints on control step $\Delta f$. Trotter steps $N_t$ and the constraints on control step $\Delta f$ determine the depth of the circuit and the continuity of the control function, respectively. In Fig. \ref{figure4}, we present the fidelity density as a function of $N_t$ and $\Delta f$ in (a) and show the maximum-fidelity control protocols for various $N_t$ and $\Delta f$, where the counter curves corresponding to the fidelity $F=0.999, 0.99$ are marked. In addition, in Fig. \ref{figure4}(b), the results from circuit learning are compared with 
	the the optimal-time protocol $t_f = t_f^{\text{opt}} $ produced by bang-bang control (\ref{bangbang}). The higher accuracy of DQS requires the larger Trotter step $N_t$ and higher computation complexity.
	On the other hand, the control function becomes smoother when $\Delta f \to 0$, with increasing the Trotter step and circuit complexity. In this context, we choose $N_t = 50$ and $\Delta f = 1$ in the following calculations.
	
	\subsection{Control phase transition at quantum speed limit}
In the previous section, we described an efficient GD-based hybrid algorithm to find the maximum-fidelity control in a quantum device by considering the loss function $J$, Trotter step $N_t$, and step length $\Delta f$.
Next, we shall prove the control phase transition (CPT)  appears at the critical point due to the QSL.
The use of optimal control to reach the QSL in quantum state manipulation has been discussed in \cite{qslwithOC}.
However,  the authors in \cite{qslwithOC} intended to find maximum fidelity control by reducing infidelity, which is a highly time-consuming task. This can be improved by introducing the fidelity susceptibility $\chi$, as we have shown in the previous section.
In the space of protocols, the control phase transition is associated with abrupt changes in an optimal control function, satisfying given constraints as the duration of the process is varied  \cite{CPT2018prx}.
In particular, the maximum-fidelity control function is unique when $t\le t^{\rm opt}$.
It is expected that the QSL can be reached at the point of CPT.

 \begin{figure}
 	\centering
 	\includegraphics[width=\columnwidth]{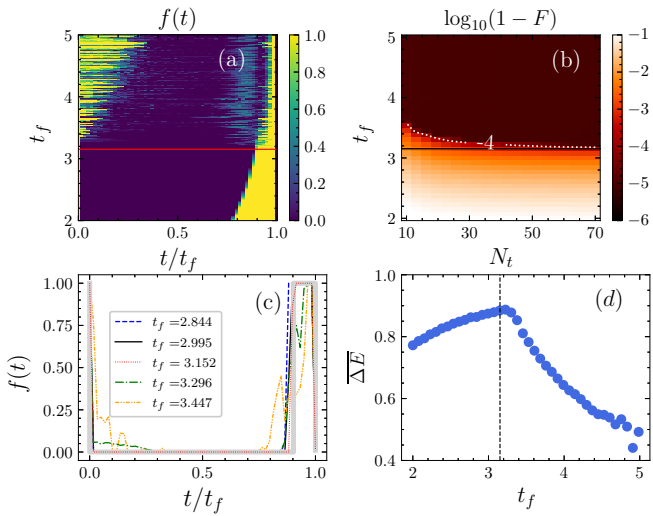}
 	\caption{ 
 	(a) The control phase diagram, with the fidelity-optimal control sequence $f(t)=\omega^2(t)$ as a function of $t/t_f$ for different $t_f\in[2,5]$.
 			(b) Logarithm of the infidelity $\log_{10}(1-F)$ as a function of ${t_f,N_t}$.  The solid line is included as a reference in (a) and (b) and corresponds to the optimal bang-bang control with $t_f^{\rm opt} =3.152$. (c)  The trained fidelity-optimal controls for different values of the total time $t_f$ are compared with the optimal bang-bang control,  (thick gray line). (d) The corresponding time-averaged energy dispersion  $\overline{\Delta E}$ for maximum-fidelity control. Parameters agree with those in Fig. \ref{figure3}.
 		}
 	\label{figure5}
 \end{figure}

In practice, we consider the expansion process:  from initial state $\psi_0$ ($\omega_0 = 1$) to target states $\psi_{\text{tar}}$ ($\omega_f = 0.1$) with constrain $\delta_1 =10^{-6}$, $\delta_2 = 1$ by assuming $\omega^2 (t) >0$, and subsequently generate the maximum-fidelity control sequence $f(t)$ for $t_f\in[2,5]$, where other parameters are same as those in Fig. \ref{figure3}.
Consequently, we present the control phase diagram in Fig. \ref{figure5}(a) and illustrate several control functions of selected $t_f$ compared with the optimal bang-bang control  (\ref{bangbang}).
The control function is suddenly converted to a non-bang-bang-typed phase at the transition point $t_f \approx t^{\text{opt}}_f$, i.e., the control phase transition point.
In addition, we note that there exists only one solution of the maximum-fidelity control function when $t_f \leq t^{\text{opt}}_f$ for fulfilling the requirement of time-energy bound (\ref{tau_QSL}).
Here, we confirm the minimum time $t_f^{\min}$ when the maximum-fidelity is larger than 0.999, and compare the result for different $N_t$ in (b) of Fig. \ref{figure5}.
It is evident in Fig. \ref{figure5}(c) that the accuracy of optimal time produced by circuit learning essentially depends on the $N_t$, and we set $N_t=50$ for the criteria of $\log_{10}(1-F)\sim -4$ in the following calculations.

We emphasize that the time-optimal driving obtained here differs from the QSL but is closely related to it. Specifically, the time-optimal driving is bounded in terms of the trap frequency by contrast to the QSL, which is bounded in terms of the time-averaged standard deviation of the energy, see Eq. (\ref{qsl-ho}). The former is a weaker and more conservative bound, as the energy fluctuations can be upper-bounded in terms of the frequency.
In Fig. \ref{figure5}(d), we display the energy dispersion of maximum-fidelity control and compare it with the corresponding time-optimal driving. 
The energy dispersion becomes consistently unique when $t_f \leq t^{\rm opt}_f$ before the point of CPT. Moreover, the energy dispersion for the maximum fidelity control is slightly smaller than in the time-optimal bang-bang control with bounded trap frequency. 
In this sense, one can approach the QSL in time-optimal driving when an additional energy cost is allowed by relaxing the trap frequency bound. 
	
\section{Discussion}

	A significant source of discussion is the robustness of VQAs in an environment with stochastic perturbations.
	In a real quantum computer with NISQ hardware,  imperfections are unavoidably induced as a result of a finite number of measurements and a noisy environment.
	As emphasized,  the previous results are produced by the quantum simulator {\tt statevector simulator} in the {\tt qiskit} platform, with no errors, decoherence, and imperfections at all.
	In this section, we implement our method in a noise-associated quantum device simulated by {\tt qasm simulator} with $N_m$ measurement shots.	
	The performance of the GD algorithm depends on the tolerance of the optimizer to errors. Thus, we shall balance the GD induced by noise and the parameter variance in a training landscape.
	Let us recall the definition of noise in the framework of quantum information processing.
	In general, the  $n$-qubits register is coupled with an environment $\varepsilon$, leading to the nonunitary evolution of the system. 
	Initially, we assume the density operators of the register $\rho(t_0)=\rho_0$ and the environment to be decoupled so that the composite state is given by the tensor product  $\rho\otimes \varepsilon$.
	For any global unitary operator $U$ describing the dynamics of the composite state, the reduced evolution of the register reads
	\beq
	\rho(t) = {\rm Tr}[U(\rho(t_0)\otimes \varepsilon)U^{\dagger}]\equiv \xi(\rho_0).
	\eeq
	This superoperator $\xi(\cdot)$ can be implemented for simulating a noise model in a quantum circuit. 
	The noisy quantum channel describes the nonunitary evolution of the time-varying density state in the Kraus representation
	\beq
	\rho(t) = \sum_kE_k\rho(t_0) E_k^{\dagger},
	\eeq
	where $E_k$  satisfy the trace-preserving condition $\sum_kE_kE_k^{\dagger} = \textbf{1}$.
	Since we perform the measurement on the qubits register only at $t=t_f$,   imperfections induced by any kind of noise result in fluctuations of the measurement accuracy.
	In this sense, we shall primarily consider the bit-flip error of measurements in a real quantum computer.
	Assume the system's noise flips $|0\rangle$ and $|1\rangle$ with probability $\beta$. The superoperator for this bit flip noise can be expressed as
	\beq
	\mathcal{\xi}_{BF}(\rho) = (1-\beta)\rho+\beta X \rho X ,
	\eeq
	where the corresponding Kraus operators are $\{\sqrt{1-\beta}\mathbb{I}, \sqrt{\beta}X\}$, in terms of the identity $\mathbb{I}$ and the Pauli operator $X$.
		\begin{figure}
	\centering
	\includegraphics[width=\columnwidth  ]{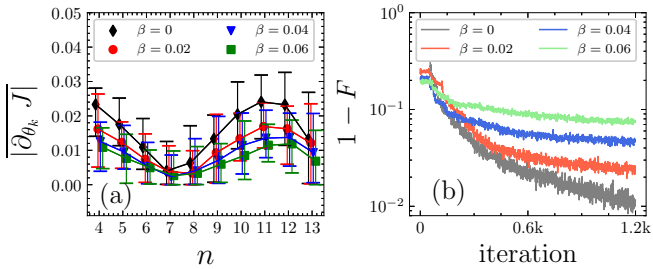}
	\caption{ (a) The average gradient $\overline{|\partial_{\theta_k}J|} $  as a function of the qubit number $n$ and number of layers given by the polynomial function $N_t = 5\times n$.
		(b) Infidelities as a function of the training iteration for values of the noise strength $\beta = 0,0.02,0.04,0.06$ based on the $\tt SPSA$ optimizer.
		Other parameters are equal to those in Fig. \ref{figure3}.
	}\label{figure6}
\end{figure}

Next, we discuss whether the \textit{barren plateau} \cite{McClean2018nc} phenomenon occurs here. The barren plateau refers to the fact that the gradient of an observable vanishes exponentially as a function of qubits number in a training landscape of VQAs. It has been widely studied in various ansatze of deep circuits \cite{VQA2021NP}. In general, the gradient of an objective function is calculated by mean of the parameter-shift rule \cite{shift2018pra,analyticgrad2019pra}, expressed as $\partial_{\theta_k} J = \frac{1}{2}[J({\theta_k}+\frac{\pi}{2})- J({\theta_k}-\frac{\pi}{2})] $ for an arbitrary trainable parameter $\theta_k$ in the circuit.
	In this sense, we define the average of the absolute gradient over $N_r$ random initializations
	\beq\label{average_gradient}
	\overline{|\partial_{\theta_k}J|} =\sum_{i=1}^{N_r} \frac{1}{2N_r} \left| J_i({\theta_k}+\frac{\pi}{2})- J_i({\theta_k}-\frac{\pi}{2})\right|.
	\eeq	 
	
Since these three objective functions we proposed in the previous section are all involve the fidelity, we here provide numerical analysis of the average gradients of the fidelity $F$.
Essentially, the probability distribution of qubit states obeys a statistical precision of order $1/\sqrt{N_m}$, and the fidelity defined in Eq. (\ref{fidelity}) meets the same criteria.
{The derivative of an observable with respect to an arbitrary trainable $\theta_k$  in the circuit is a linear function of the gradient with respect to the corresponding control parameter $f_k = f(k)$ at  $k$-th Trotter step: $\partial _{f_{k}} J = 2c\cdot\partial _{\theta_{k}} J$, with $c$ being a real number. Thus, the gradient of the objective function with respect to the control parameter $f(t)$ shares the analytic expression in Eq. (\ref{average_gradient}), see also the detail in Appendix \ref{App1}.}	 
Also, we calculate the average gradient Eq. (\ref{average_gradient}) over $N_r = 50$ random initialization of $f(t)$ for various qubit number $n$, while the Trotter step is taken as a polynomial function as $N_t = ploy(n)$. In Fig. \ref{figure6}(a), we demonstrate that the \textit{barren plateau}  is avoided for the average gradient of a correlating parameter $\theta_k$.
The main reason for the absence of \textit{barren plateau} is the reduction of ansatz's expressibility, due to the strong correlation of parameter $\theta_k$ depending on the controller $f(t)$ in our method, a common feature with the recent work in Ref. \cite{Expressibility2022prxQ}.
In this regard, we set measurement shots $N_m = 8192$ for statistical accuracy and energy saving by considering the gradient magnitudes, as shown in Fig. \ref{figure6}(a).
Moreover, we apply the optimizer of simultaneous perturbation stochastic approximation ({\tt SPSA}), which is widely used for solving an optimization problem with statistical noise \cite{SPSA,MLbook}. In Fig. \ref{figure6}(b), we present the infidelity as a function of the training iteration for $\beta = 0,0.02,0.04,0.06$, where the infidelity for noise-free case ($\beta = 0$) convergent to $\sim10^{-2}$ which obeys the criteria of $\sim 1/\sqrt{N_m}$.
Moreover, the performance of {\tt SPSA} is compared with various optimizers in Appendix \ref{App2}.

Let us discuss the circuit complexity in terms of the qubits number $n$ and the number of Trotter steps $N_t$. The whole circuit consists of  $N_t$ circuit units that simulate each unitary operation $ \hat{U}(dt) =e^{-iHdt}$, as depicted in Fig. \ref{figure1}.
In the absence of $\theta_1$ and $\theta_2$, the gate number of each circuit unit, including the potential operator  $\mathcal{V}(dt)$ and  the kinetic-energy operator $\mathcal{T}(dt)$ in real space and momentum space, is $N_{unit}\sim 2n^2$.
In addition, the operation of  \text {QFT} and \text {iQFT} requires $\sim n^2/2$ control-phase gates. Consequently, the total number of gates for our ansatz is proportional to a quadratic function of $n$, yielding $ 5N_t n^2/2$. 
To find the minimal-time control on the QC with reasonable precision, one can increase the qubit number $n$ and Trotter step $N_t$, with exponentially enlarged Hilbert space. But this leads to the quadratic size increase in circuit complexity. Recently, alternative methods inspired by the Grover-Rudolph algorithm \cite{marin2021quantum} are worked out for the quantum state preparation, which are expected to reduce its complexity in this direction.

Finally, we discuss the control problem beyond the quadratic Hamiltonian on which we have focused. For our case study, one can introduce perturbations of the trap, e.g., a time-independent anharmonicity involving an operator $x^4$, which is no longer quadratic. Although its exact decomposition into quantum circuits does not exist, one can still approximate the evolution block with arbitrary precision by the Solovay-Kitaev algorithm \cite{kitaev1997quantum}, placing it in the block of $\mathcal{V}_{dt/2}$ before or after the evolution of the quadratic Hamiltonian since it commutes with the harmonic potential operator.

	\section{Conclusion}

To sum up, we propose the GD-based circuit learning to find the time-optimal control problem, the driving of a quantum particle trapped in a time-varying harmonic potential, and figure out its quantum speed limit in relation to the control phase transition.  
First, we have constructed the digitized quantum circuit of a  time-dependent harmonic oscillator using a finite $n$-qubit register. Second, we have demonstrated that the learning rate of circuit optimization can be accelerated by considering various physical quantities, such as the infidelity, Bures angle, and fidelity susceptibility, as cost functions, thus reducing training iteration.
	Third, we have established the relation between control phase transition and quantum speed limit.
	Finally, we have established the error tolerance of our method by considering the presence of measurement errors in a  quantum computer. 
	The absence of a barren plateau is further justified in our ansatz, enabling the application of VQAs for a class of tasks that is not affected by the fundamental limitations of NISQ devices.
	As a heuristic example, we have demonstrated that quantum control can be efficiently simulated and optimized using a NISQ device by combining digital quantum simulation and hybrid circuit learning. Numerical experiments prove that barren plateaus are avoided in the framework.

\section*{ACKNOWLEDGMENTS}
This work was financially supported by NSFC (12075145), STCSM (Grants No. 2019SHZDZX01-ZX04), EU FET Open Grant  EPIQUS (899368),  the Basque Government through Grant No. IT1470-22, the project grant  PID2021-126273NB-I00 funded by MCIN/AEI/10.13039/501100011033 and by ``ERDF A way of making Europe" and ``ERDF Invest in
your Future", QUANTEK project (KK-2021/00070), and the BRTAQ project (expendient no. KK-2022/00041). HTY acknowledges CSC fellowship (202006890071). X.C. acknowledges ayudas para contratos Ramón y Cajal--2015-2020 (RYC-2017-22482).

\appendix

\section{ The parameter-shift rule}\label{App1}

\begin{figure}[t!]
	\centering
	\includegraphics[width=\columnwidth]{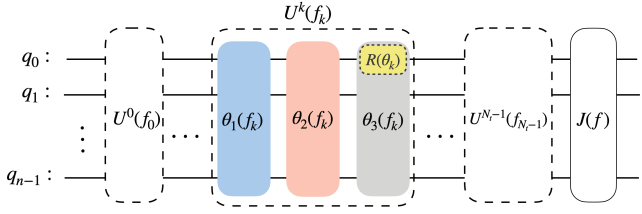}
	\caption{Schematic illustration of a quantum circuit with Trotter step $N_t$. Each dashed-block denoted in Fig. \ref{figure1})
	presents the implementation of time-evolution $e^{-iH(t)dt}$ with quadratic Hamiltonian $H(t)$ (\ref{quadratic operator}). Here, an arbitrary operator $U^{k}(f_k)$ includes three $f_k$-correlated variables: $\theta_1(f_k)$, $\theta_2(f_k)$ and $\theta_3(f_k)$. The single-qubit gate with a yellow shadow is used to calculate the gradient of the objective function.
	}\label{figure_app1}
\end{figure}

One recipe to find the partial derivative of an objective function $J(\Theta)$ in parametric quantum circuits (PQCs) is known as the parameter-shift rule \cite{shift2018pra,analyticgrad2019pra}.
	In general, the expectation value of an observable $\hat{B}$ as a function of a single parameter $\theta_k$ in a circuit is expressed as $J(\theta_k) =\langle \hat{B}(\theta_k) \rangle$.
	We assume a sequence of unitary operations represented as $U(\theta_k) = U_LU_k(\theta_k)U_R$, for which we have
	\beqa\label{J_appA}
		J(\theta_k) &=& \langle0| U_R^{\dagger}  U_k^{\dagger}(\theta_k)U_L^{\dagger} \hat{B} U_LU_k(\theta_k)U_R | 0\rangle \nonumber \\
		&=&\langle z| \mathcal{M}(\hat{B},\theta_k) |z\rangle,
		\eeqa
		where $\mathcal{M}(\hat{B},\theta_k) =U_k^{\dagger}(\theta_k)U_L^{\dagger} \hat{B} U_L U_k(\theta_k) $ and the basis reads $|z\rangle = U_R | 0\rangle$.
		Consider a unitary operator $U_k(\theta_k)$  generated by a Pauli matrix $\sigma_k $ as $U_k(\theta_k) =  \exp(-i\theta_k\sigma_k/2)$.
		The gradient of the objective function is defined as
		\beqa\label{define_gad}
		\partial_{\theta_k} J(z;\theta_k)  &=&  \langle z| \partial_{\theta_k} \mathcal{M}(\hat{B},\theta_k) |z\rangle \nonumber \\
		&=& c
		\left[\langle z| \mathcal{M}(\theta_k+s) |z\rangle -\langle z| \mathcal{M}(\theta_k-s) |z\rangle\right], ~~~
		\eeqa
		where coefficient $c$ and shift $s$ are independent of $\theta_k$.
		The gradient $\partial_{\theta_k} U_k(\theta_k) = -\frac{i}{2}U_k(\theta_k)\sigma_k$, and inserting it to (\ref{J_appA}), we have \cite{McClean2018nc}
		$ \partial_{\theta_k} J = -\frac{i}{2}\langle z|U_k^{\dagger}(\theta_k)[\sigma_k,\hat{B}]U_k(\theta_k)|z\rangle$.
		The commutation relation,
		$
		[\sigma_k,\hat{B}] = i \left(U_k^\dagger(\frac{\pi}{2})\hat{B}U_k(\frac{\pi}{2})-U_k^\dagger(-\frac{\pi}{2})\hat{B}U_k(-\frac{\pi}{2})\right)$, 
		 enables us derive the analytical gradient as \cite{shift2018pra,analyticgrad2019pra}:
		\beqa\label{shift_rule}
		\partial_{\theta_k}J&=& 
		\frac{1}{2}
		\langle z|\left[U_k(\theta_k^+)\hat{B}U_k(\theta_k^+)-U_k^\dagger(\theta_k^-)\hat{B}U_k(\theta_k^-)\right]|z\rangle \nonumber \\
		&=& \frac{1}{2}
		\left[\langle z| \mathcal{M}(\theta_k^+) |z\rangle -\langle z| \mathcal{M}(\theta_k^-) |z\rangle\right],\nonumber 
		\eeqa
		with $\theta_k^{\pm} = \theta_k \pm \pi/2$.
		The above expression provides an analytical evaluation of the gradient of an objective function involving Pauli operators.

		In Fig. \ref{figure_app1}, we schematically illustrate the deep circuit of our method, which is composed of $N_t$ dashed blocks (refers to $N_t$ Trotter steps). In each unitary operator $U^{k}(f_{k})$, all rotating parameters $(\theta_1,\theta_2,\theta_3)$ are $f_{k}$-correlated, as detailed in Fig. \ref{figure1}.
		Let us now select an arbitrary single-qubit gate $R(\theta_k)$, with a gradient obeying the parameter shift rule in Eq. (\ref{shift_rule}). 	
	According to the algorithm in Fig. \ref{figure1}, the rotation $\theta_k$ is correlated with $f_{k}$ as a liner form: $\theta_k =  c_1 \cdot f_{k}$, where $c_1$ is a real number. Starting with Eq. (\ref{define_gad}), we have 
	    $ \partial _{f_{k}} J  = c [J(f_{k}+s)-J(f_{k}-s)]
		= c  [J(\theta'_k+s)-J(\theta'_k-s)]$, 
		where the gradient is independent of the initial angle $\theta'_k = \theta_k/c_1$. 
		Consequently, we find $
		\partial _{f_{k}} J = 2c\cdot\partial _{\theta_{k}} J$,  while the shift $f_{k}$ is $s = \pi/2$. 
		Furthermore, we introduce the notation
		\beq
		\overline{|c|} =\sum_{i=1}^{N_r} \frac{1}{2N_r} \left| \frac{\partial _{f_k}J }{\partial _{\theta_{k}} J}\right|,
		\eeq
		where the absolute average value $\overline{|c|} $ over $N_r$ random initialization.
		In Fig. \ref{figure_app2}(a), we calculate the average value $\overline{|c|} $ with $N_r = 50$, which is irrelevant to $n$.

		\section{Comparisons of optimizers}\label{App2}
		
\begin{figure}[b]
			\centering
\includegraphics[width=\columnwidth ]{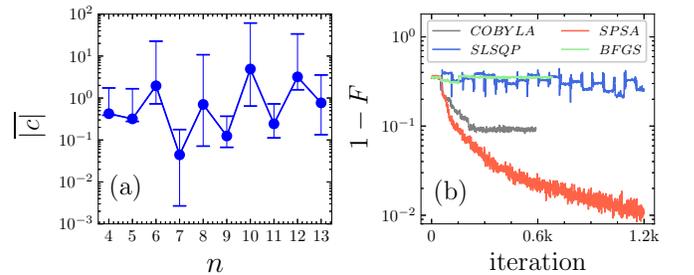}
				\caption{(a) The coefficient $\overline{|c|}$ as a function of qubit number $n$. (b) Various training processes for using different optimizers are illustrated. All calculations are proceeded with the parameters $\beta = 0$, $N_m =8192$, and others in Fig. \ref{figure3}.
				}\label{figure_app2}
	\end{figure}

		In Appendix (\ref{App2}), we compare the performance of several classical optimizers for the same optimization task with statistical errors from a finite number of measurements.
		We choose widely used optimizers, namely, $\tt SLSQP$, $\tt COBYLA$ ,$\tt SPSA$ , $\tt BFQS$ based on the library of $\tt qiskit$.
		In Fig. \ref{figure_app2}(b), we present the infidelity as a function of the training iteration by using various optimizers. The {\tt SPSA} stands out for its performance in an optimization task 
		in the presence of bit-flip noise.

	\bibliographystyle{apsrev4-1}
	\bibliography{ref}	
	
\end{document}